\documentclass[float,aps,preprint,prc]{revtex4}
\usepackage{amsmath}
\usepackage{epsfig}
\usepackage{color}
\usepackage{endnotes}
\let\footnote=\endnote

\newcommand{\be}{\begin{equation}}
\newcommand{\ee}{\end{equation}}

\newcommand{\ber}{\begin{eqnarray}}
\newcommand{\eer}{\end{eqnarray}}

\begin{document}

\title{Integral transform methods:
a critical review of various kernels} 
\author{Giuseppina Orlandini$^{1,2}$, Francesco Turro$^{1}$
  }

\affiliation{
  $^{1}$Dipartimento di Fisica, Universit\`a di Trento, I-38123 Trento, Italy \\
  $^{2}$Istituto Nazionale di Fisica Nucleare, Gruppo Collegato di Trento,
  I-38123 Trento, Italy 
}

\begin{abstract}
Some general remarks about integral transform approaches to response functions are made. Their advantage for calculating cross sections 
at energies in the continuum is stressed. In particular we discuss the class of kernels that allow calculations of the transform by 
matrix diagonalization. A particular set of such kernels, namely the wavelets, is tested in a model study.  
\end{abstract}
\maketitle
\section{Introduction}
\label{intro}
Integral transform methods are a very useful tool in all fields of physics, in experiment as well as in theory. 
In fact one often encounters situations where  an observable is not directly accessible, while this is the case for  
one of its integral transforms. If one denotes as $f(x)$ the function representing an observable of 
interest, its integral transform is defined as  
\begin{equation}
g(\sigma_1, \sigma_2,...)=\int dx f(x) K(x,\sigma_1, \sigma_2,... )\,.
\end{equation}
Depending on the  kernel $ K$ it may happen that, while one is not able to access $f(x)$, there is some reason 
why it is much easier to have access to $g(\sigma_1, \sigma_2,...)$. For example in an experiment the {\it window}
of a detector may represent the kernel 
$K$, since the observer does not have a direct access to the observable itself, but to the result of its folding 
with that particular {\it window}.
If one wants to know the original observable one has to {\it unfold} the data, namely one has to invert the transform.  

In theoretical physics there are many examples of integral transform applications, ranging from QCD, to nuclear physics, 
to condensed matter physics. In most cases $f(x)$ is a quantum mechanical quantity involving {\it continuum states}. 
It is well known that such states are the most difficult objects to access theoretically. For example in non relativistic 
quantum many-body systems, knowing a continuum state implies to solve the many-body body scattering problem, which has a viable solution 
in  a very limited number of cases. Therefore scattering cross sections or spectra involving more than three or four particles are often 
inaccessible in a direct way, namely by first calculating the wave functions entering the dynamical matrix elements 
between initial and final states. However, for some particular kernels,
the calculation of an integral transform of such cross sections/spectral functions is possible. 
One then remains with the task of extracting as much information 
as possible about $f(x)$, namely one is left with the inversion problem.

The inversion of integral transforms, when they are affected by errors of any nature, suffers from the well known {\it ill-posed problem},
namely the inversion may generate even very different functions whose transforms lie within the error uncertainty. 
Various algorithms deal with this problem. Among them the so called {\it regularization methods}, as well as  methods based 
on  Bayes' theorem, like the Maximum Entropy Method. The success of an inversion algorithm, however, also depends on the form 
of the kernel. For example, 
kernels that are 
representations of the $\delta$-function, like in the Lorentz integral transform (LIT) method~\cite{Ref1}, 
are very powerful. The reason is that the width of the 
$\delta$-function representation allows to concentrate as much information as possible 
in the transform, facilitating in this way the task of recovering that information in the inversion process.

The LIT method~\cite{Ref1} has been largely and successfully applied to electro-weak interactions with few-nucleon 
systems. In view of an evolution and an extension of this method,  we would like to discuss here
which kinds of kernels satisfy the two essential conditions for getting sufficient knowledge on $f(x)$,
i.e. i) the calculability of the transform and ii) the stability of its inversion.
(Such a discussion applies equally well to any other few/many-body system of different nature).

\section{Integral transforms via diagonalization methods}\label{sec:1}
As was already said above using integral transform approaches is particularly advantageous when the calculation of an 
observable of interest requires the knowledge of continuum states. Since several 
accurate methods have been devised for bound-states of few/many-body systems (see~\cite{Ref2} for a review) it is interesting 
to ask which are the conditions that a kernel has to fulfill in order to allow the transform to be calculated by bound state methods.
Among them we will focus in particular on the methods based on the diagonalization of the Hamiltonian  represented 
on finite norm basis functions. 

We start identifying $f(x)$ with  the excitation spectrum/response function $S(\omega)$ of a system  perturbed 
by an external probe:
 \begin{equation}\label{S}
 S(\omega)=\sum_n|<n|{\cal O}|0>|^2\delta(\omega-E_n+E_0)\,,
 \end{equation}
where ${\cal O}$ and $H$ are excitation operator and Hamiltonian of the system, respectively ($H|n>=E_n|n>$).
An integral transform of $S(\omega)$ with a kernel $K(\omega,\sigma)$ is ($\sigma$ may indicate one or more parameters)
\begin{equation}
 \Phi(\sigma)=\int \,d\omega \,S(\omega)\, K(\omega,\sigma)\,.\label{trasf}
 \end{equation}
 Substituting  $S(\omega)$ in~(\ref{S}), integrating in $d\omega$ and using the Schr\"odinger equation 
 and the completeness property of the Hamiltonian eigenstates, 
 (\noindent$\sum_n |n><n|=I$) the transform can be expressed as
 \begin{equation}
 \Phi(\sigma)= <0|{\cal O}^\dagger\,K(H-E_0,\sigma)\,{\cal O}|0>\, \,.
 \end{equation}
If the kernel is a positive definite function of $\omega$ one can split it into  $k^{\dagger}k$, therefore
 \begin{equation}
 \Phi(\sigma)= <0|{\cal O}^\dagger\,k^{\dagger}(H-E_0,\sigma)k(H-E_0,\sigma)\,{\cal O}|0>\equiv<\tilde\Psi|\tilde\Psi>\, \,.
 \end{equation}
If the transform $\Phi(\sigma)$ exists, namely the integral in (\ref{trasf}) is finite and ${\cal O}|0>$ has 
a finite norm (which is the case for all physical excitation operators) 
one can expand $|\tilde\Psi>$ e $<\tilde\Psi|$, as well as ${\cal O}|0>$ and $<0|{\cal O}^\dagger$, on a complete set  of finite 
norm functions
  ($|l>, |m>, |p>$). Therefore
 \begin{equation}
 \Phi(\sigma)= \sum_{l,m,p} \, <0|{\cal O}^\dagger|m><m|\, k^{\dagger}(H-E_0,\sigma)|l><l|k(H-E_0,\sigma) |p><p|\,{\cal O}|0> \,.
 \end{equation}
In this way the kernel is a  function of the matrix elements $H_{ml}=<m|H|l>$ or $H_{lp}=<l|H|p>$  of the Hamiltonian 
represented on those basis functions. After diagonalizing the matrix, 
$\Phi(\sigma)$ becomes:
 \begin{equation}
 \Phi(\sigma)=\sum_{\mu} <0|{\cal O}^\dagger\,|\mu><\mu|\,k^{\dagger}(\epsilon_\mu-E_0),\sigma)|\mu>
 <\mu|k(\epsilon_\mu-E_0, \sigma),\sigma|\mu><\mu|{\cal O}|0>\, \,.
 \end{equation}
Recombining $k^{\dagger}(\epsilon_\mu-E_0,\sigma )k(\epsilon_\mu-E_0,\sigma)$ into  $K(\epsilon_\mu-E_0,\sigma)$ one has 
 \begin{equation}
 \Phi(\sigma)=\sum_{\mu} K(\epsilon_\mu-E_0,\sigma) |<\mu|{\cal O}|0>|^2\, \,.\label{theorem}
 \end{equation}

Summarizing: it is possible to calculate integral transforms of spectra/response functions, also to continuum,  via diagonalization 
of the Hamiltonian represented on finite norm basis functions, provided that the following conditions are satisfied:
i) the integral transform exists, namely $\Phi(\sigma)\, <\, \infty $;
ii) the $0^{th}$ moment of $S(\omega)$ exists, namely 
$\Phi(\sigma)=\int d\omega S(\omega)\, <\, \infty\,  (\,\,\Longrightarrow \Phi(\sigma)\, = \, <0|{\cal O}^\dagger\,{\cal O}|0>\,\,)\,;
$
iii) the kernel is a positive definite function of ${\omega}$ or a linear combination of them.
 
The convenience of the kernel depends primarily on condition i) 
(e.g. the  moment kernel $K(\sigma,\omega)=\omega^\sigma$, with $\sigma$ integer, may  fulfill that condition only 
for a very limited number of $\sigma$  values), 
but also by the possibility to invert the transform reliably. In fact there are some kernels for which the inversion of the transform is a 
seriously {\it ill-posed problem}. A well known example is the  Laplace kernel  $exp[- \sigma \omega]$,
widely used in imaginary-time-Monte-Carlo calculations.
Another example is $K(\sigma,\omega)=(\omega-\sigma)^{-1}$, suggested in~\cite{Ref3} and known as the Stieltjes kernel, which, however, 
has turned out to be very useful for direct calculations of polarizabilities~\cite{Ref4}, 
quantities that also contain the physics of the continuum.

\section{A special class of kernels: wavelets}

A special class of kernels, satisfying  the iii) condition expressed after Eq.~(\ref{theorem}), is represented 
by the {\it wavelets}~\cite{Ref5}.
These kernels combine the advantages of the Fourier analysis with the property of focusing on different ranges of 
$\omega$, since they are defined on finite intervals.
In this respect they present the same advantage as the Lorentzian kernel, whose two parameters rule the position and width of the 
{\it window}-like kernel.
Moreover, some of them form a complete set of orthogonal functions, allowing a straightforward and controllable 
inversion expression for the transform.

\begin{figure*}
\centering
\includegraphics[width=0.4\textwidth]{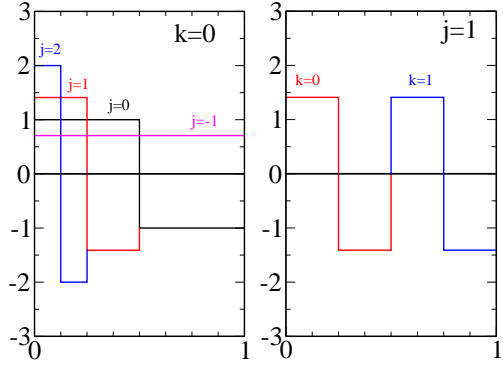}
\caption{Left: $K_W(\omega,k,j)$ at fixed $k$ for various $j$. 
Right: $K_W(\omega,k,j)$ at fixed $j$ for various $k$}
\label{fig:1}       
\end{figure*}

In order to get familiar with these kernels, and in view of the possibility to use them in physical cases, we have chosen to investigate 
the performances of a particularly simple class of them, i.e. the Haar wavelets. They have been the first kind of wavelets 
to be introduced in 1909 by 
Alfred Haar~\cite{Ref6} as an example of a numerable orthonormal system of ${\cal L}_2$ functions.
They are generated by the {\it mother wavelet} defined by:
\begin{equation}
\psi(z)=\left\{ \begin{array}{cc}
1 & if \; 0<z<\frac{1}{2}\\
-1 & if \;  \frac{1}{2}<z<1\\
0 & elsewhere
\end{array}
\right. ;
\end{equation} 
Therefore, denoting with $\Omega$ the energy range of interest, 
the wavelet kernel is given by 
\begin{equation}
K_W(\omega,k,j)=2^{\frac{j}{2}} \left[
     \Theta\,\left(\omega- \frac{k}{2^{j}}\Omega\right)+
     \Theta\,\left(\omega - \frac{k+1}{2^{j}}\Omega\right) - \,2\,\Theta\,\left(\omega-\frac{2 k+1}{2^{j+1}}\Omega\right)\, \right]\,, 
\end{equation}
where $\Theta$ is the Heaviside function and $k$ and $j$ are integer numbers determining the position and width of the wavelet,
respectively ($j=-\infty,...,-1,0,1,...,\infty$ and $k=0,1,2...k_{max}$ with $k_{max}=2^{j}-1$). 
One can easily verify that the Haar wavelets fulfill the zero-mean condition and are orthonormal. 
Some of them are plotted in Fig.~\ref{fig:1}.
With such a kernel $\Phi(k,j)$ becomes a function of integer variables,
and because of the orthonormality relation of the wavelets, the inversion is simply obtained by
 \begin{equation}
 S(\omega)=\sum_{j}^\infty\sum_k^{k_{max}} \Phi(k,j) K_W(\omega,k,j)\,.\label{inv}
\end{equation}

\begin{figure*}
\includegraphics[width=0.42\textwidth]{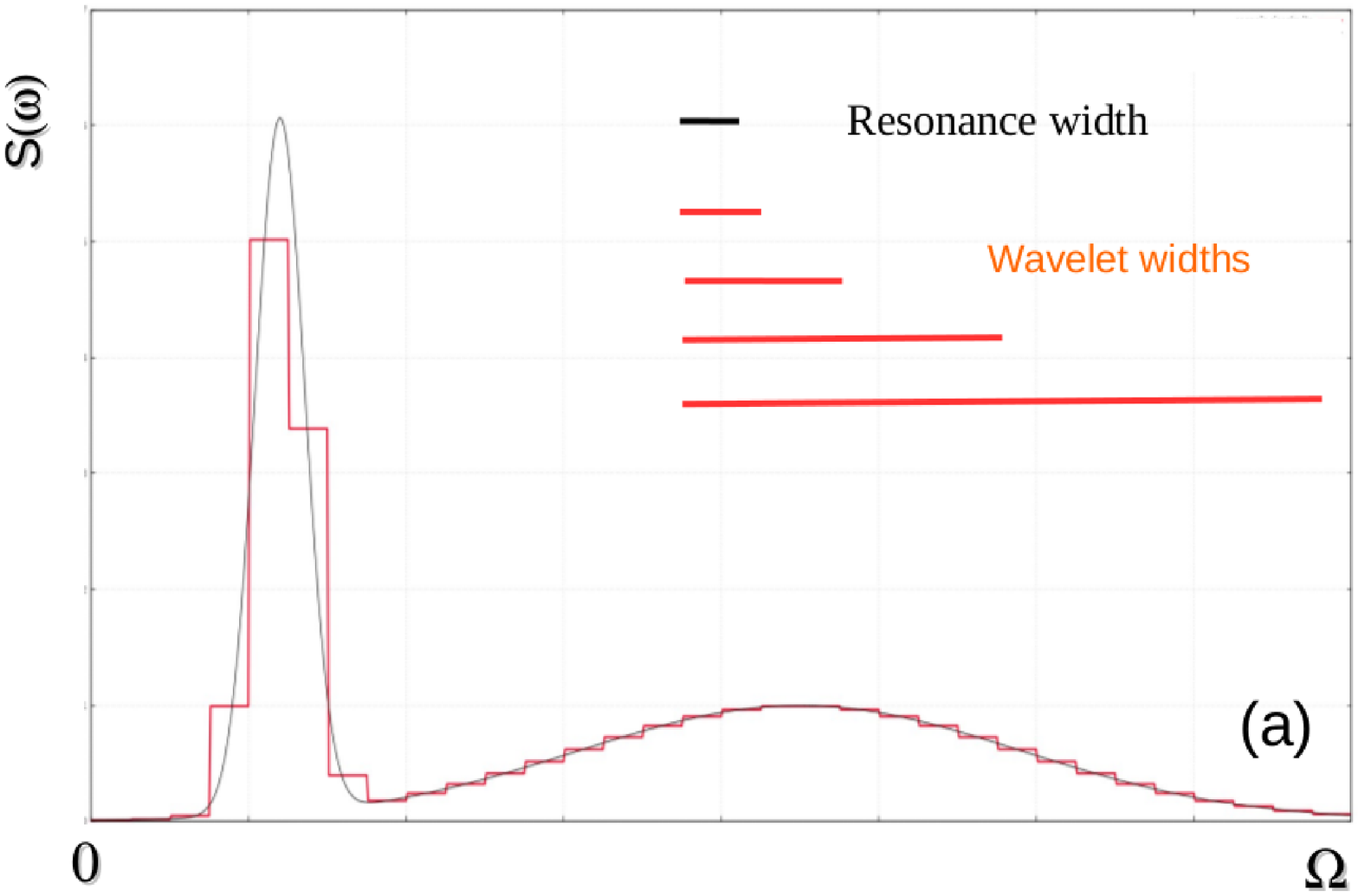}
\includegraphics[width=0.42\textwidth]{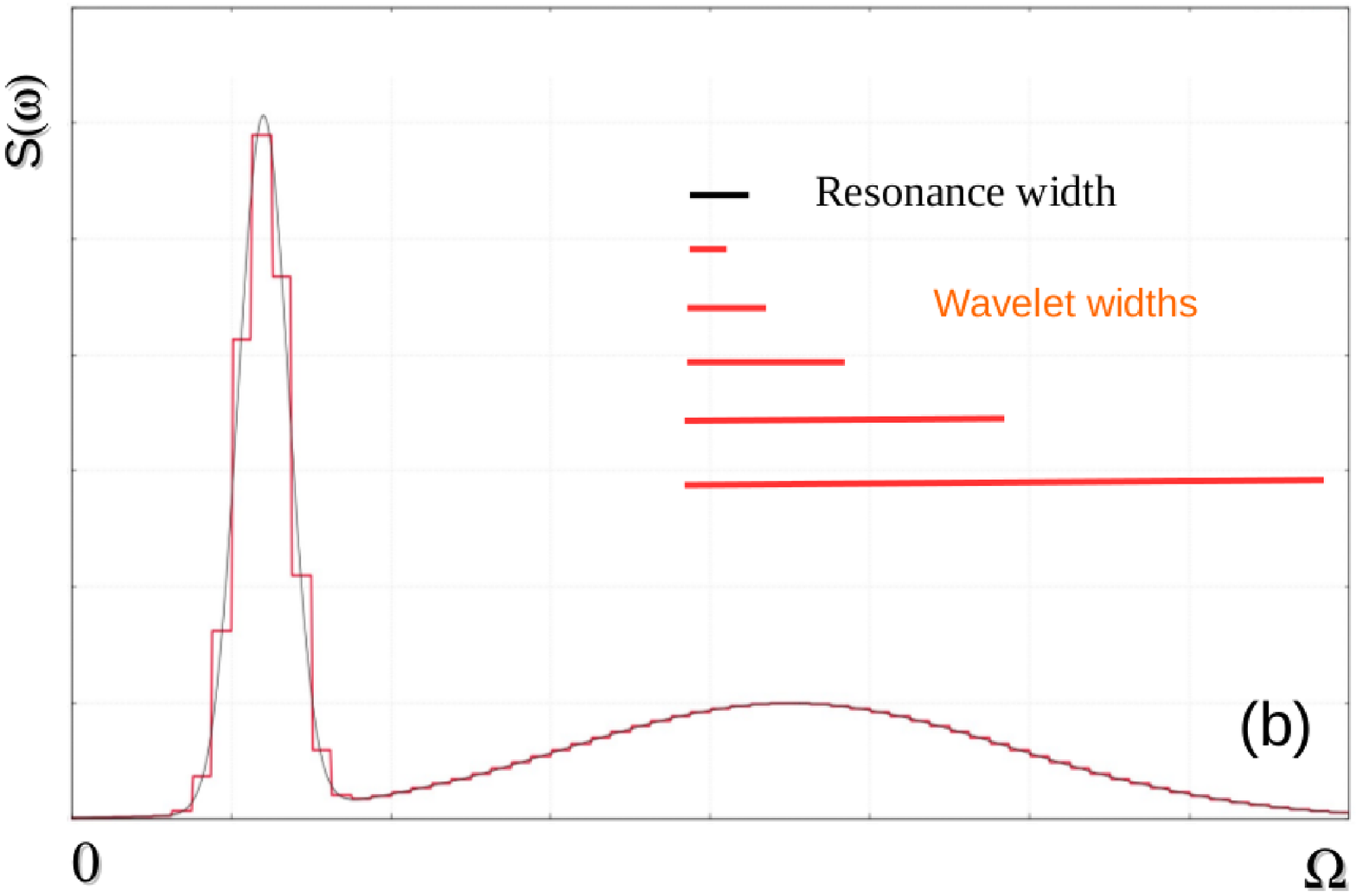}
\caption{Model $S(\omega)$ (black curve) and the result of the 
inversion of the wavelet transform (red curve) 
for $j_{max}=4$ (a) and $j_{max}=5$ (b) 
in the sum of Eq.~(\ref{inv}).}
\label{fig:2}        
\end{figure*}

In the following we show the quality of the {\it reconstruction} of a model $S(\omega)$, starting from the knowledge of 
its wavelet transform.
For the test we have chosen an $S(\omega)$ resembling a physical spectrum characterized by a narrow resonance 
and a larger peak,
a situation which is often problematic for common inversion methods. 
It is instructive to see the quality of the inversion at different values of $j_{max}$ 
in Fig.~\ref{fig:2}a and~\ref{fig:2}b. 
The values of the index $j$, in fact, represent the various {\it resolution} powers of the wavelets. Therefore fixing $j_{max}$ corresponds 
to fix the maximum resolution power of the wavelet transform.
One can notice that a maximum resolution slightly smaller than the width of the resonance allows a rather good description 
of both the resonance and the larger peak (of course a much courser resolution is enough to reproduce the larger peak only). 

We have also studied the propagation of both systematic and random errors in the inversion procedure. Eq.~(\ref{inv}), a consequence 
of the orthonormality of the wavelets, ensures that  
large instabilities cannot be generated. The steps in the reconstructed $S(\omega)$ are the consequence of using discrete wavelets. 
We intend to use also continuous wavelets, even if they have the drawback to be in general non-orthonormal.
More important will be the investigation of the realistic case, when $\Phi(k,j)$ is calculated by diagonalization. 
As Eq.~(\ref{theorem}) shows, depending on the 
basis functions used to represent $H$ (see discussion in~\cite{Ref6}) and on  the matrix size, $\Phi(k,j)$ contains the information
on the dynamics only via a discrete set of eigenvalues $\epsilon_\mu$. Their density in the range $\Omega$ determines the maximum resolution one can use. In this sense, one has again, 
as for the Lorentz kernel, a controlled resolution prediction of $S(\omega)$, namely one will be able to predict only structures
comparable or larger than the average distance among the  various $\epsilon_\mu$. The ideal case will be when 
this coincides  with the experimental resolution.

\end{document}